\documentclass[twocolumn,prb,showpacs,preprintnumbers,amsmath,amssymb]{revtex4}

\usepackage{graphicx}
\usepackage{dcolumn}
\usepackage{bm}

\begin{document}

\title{Magnetic moment and magnetic anisotropy of linear and zigzag
4{\it d} and 5{\it d} transition metal nanowires: First-principles calculations}

\author{J. C. Tung$^1$ and G. Y. Guo$^{1,2}$\footnote{E-mail: gyguo@phys.ntu.edu.tw} }
\affiliation{$^1$Department of Physics and Center for Theoretical Sciences, National Taiwan University, Taipei
106, Taiwan\\$^2$Graduate Institute of Applied Physics, National Chengchi University, Taipei 116, Taiwan}
\date{\today}

\begin{abstract}
An extensive {\it ab initio} study of the physical properties of both linear and zigzag atomic chains of 
all 4$d$ and 5$d$ transition metals (TM) within the generalized gradient approximation by using the accurate 
projector-augmented wave method, has been carried out. The atomic structures of equilibrium and metastable 
states were theoretically determined. All the TM linear chains are found to be unstable against the 
corresponding zigzag structures. All the TM chains, except Nb, Ag and La, have a stable (or metastable)
magnetic state in either the linear or zigzag or both structures. Magnetic states appear also in the 
sufficiently stretched Nb and La linear chains and in the largely compressed Y and La chains. The spin 
magnetic moments in the Mo, Tc, Ru, Rh, W, Re chains could be large ($\geq$1.0 $\mu_B$/atom). Structural 
transformation from the linear to zigzag chains could suppress the magnetism already in the linear chain, 
induce the magnetism in the zigzag structure, and also cause a change of the magnetic state (ferromagnetic 
to antiferroamgetic or vice verse). The calculations including the spin-orbit coupling reveal that the 
orbital moments in the Zr, Tc, Ru, Rh, Pd, Hf, Ta, W, Re, Os, Ir and Pt chains could be rather 
large ($\geq$0.1 $\mu_B$/atom). Importantly, large magnetic anisotropy energy ($\geq$1.0 meV/atom) is 
found in most of the magnetic TM chains, suggesting that these nanowires could have fascinating applications 
in ultrahigh density magnetic memories and hard disks. In particular, giant magnetic anisotropy energy 
($\geq$10.0 meV/atom) could appear in the Ru, Re, Rh, and Ir chains. Furthermore, the magnetic anisotropy 
energy in several elongated linear chains could be as large as 40.0 meV/atom. A spin-reorientation transition 
occurs in the Ru, Ir, Ta, Zr, La and Zr, Ru, La, Ta and Ir linear chains when they are elongated.
Remarkably, all the 5$d$ as well as Tc and Pd chains show the colossal magnetic anisotropy (i.e., it is 
impossible to rotate magnetization into certain directions). Finally, the electronic band structure and 
density of states of the nanowires have also been calculated in order to understand the electronic origin
of the large magnetic anisotropy and orbital magnetic moment as well as to estimate the conduction electron 
spin polarization.

\end{abstract}


\maketitle

\section{Introduction}

Magnetism in nanostructures
has been a very active research area in the last decades~\cite{Elmers94,Hei00,Pie00,Gambar02},
because of its novel fundamental physics and fascinating potential applications.
Experimentally, modern methods of preparing metal nanowires have made it possible
to investigate the influence of dimensionality on the magnetic properties.
For example, Gambardella, {\it et al.}\cite{Gambar02}, recently succeeded in
preparing a high density of parallel atomic chains along steps by growing Co on a
high-purity Pt (997) vicinal surface and also observed one-dimensional (1D) magnetism in a narrow
temperature range of 10$\sim$20 K. In the mean time, Li, {\it et al.}\cite{Li01} reported that Fe
stripes on the stepped Pd(110) substrate have a different magnetic easy axis than previous
results. Structurally stable nanowires can also be grown inside tubular structures, such as
the Ag nanowires of micrometer lengths grown inside self-assembled organic
(calix[4]hydroquinone) nanotubes\cite{Suh03}. Short suspended nanowires have been
produced by driving the tip of scanning tunneling microscope into contact with a
metallic surface and subsequent retraction, leading to the extrusion of a limited
number of atoms from either tip or substrate\cite{Rubio96}. Monostrand nanowires of
Co and Pd have also been prepared in mechanical break junctions, and full
spin-polarized conductance was observed\cite{Rodrigues03}.

Theoretically, a great deal of research has been done on both finite and infinite 
chains of metal atoms.  Theoretical calculations at either semi-empirical tight-binding 
or {\it ab initio} density functional theory level for
many infinite/finite chains, e.g., linear chains of Co\cite{Dallmeyer00, Jisang03, Ederer03,
 Lazarovits03, Matej02},Fe\cite{Ederer03,Spisak02}, Ni, Pd\cite{Spisak03, Lucas07}, Pt, Cu\cite{Dallmeyer00},
Ag\cite{Ribeiro03, Nautiyal03}, and Au\cite{Bahn01, Delin03, Ribeiro03, Skorodumova00, Maria00},
as well as zigzag chains of Fe\cite{Spisak02}, Zr\cite{lin04} and Au\cite{Skorodumova00},
have been reported. Early studies of infinite linear chains of
Au \cite{Portal99, Portal01, Maria00, Skorodumova00}, Cu\cite{Nautiyal03},
and Pd\cite{Delin03} have shown a wide variety of stable and metastable
structures. Recently, the magnetic properties of transition metal infinite
linear chains of Fe, Co, Ni, have been calculated \cite{Nautiyal04,Spisak02,Jisang03,Matej02,Ederer03}.
Possible magnetism in $s$- and $sp$-electron element linear and zigzag chains have
also been studied theoretically.\cite{zhu09}
These calculations show that the metallic and magnetic nanowires may become important for
electronic/optoelectronic devices, quantum devices, magnetic storage, nanoprobes and
spintronics.

Despite of the above mentioned intensive theoretical and experimental research,
current understanding on the intriguing magnetic properties of nanowires and how magnetism
depends their structural property is still incomplete. The purpose
of the present work is to make a systematic {\it ab initio} study of the magnetic,
electronic and structural properties of linear and zigzag atomic chains (Fig. 1)
of all 4$d$ and 5$d$ transition metals (TM). Transition metals, because of their partly
filled $d$ orbitals, have a strong tendency to magnetize. Nonetheless, only 3$d$
transition metals (Cr, Mn, Fe, Co, and Ni) exhibit magnetism in their bulk structures.
It is, therefore, of interest to investigate possible ferromagnetic (FM) and
antiferromagnetic (AF) magnetization in the linear chains of all 4$d$ and 5$d$ transition
metals including Y, Zr, Nb, La, Hf and Ta zigzag chain which appear not to have been considered.
As mentioned before, recent {\it ab initio} calculations indicate that the zigzag chain structure
of, at least, Zr\cite{lin04}, Ir\cite{Lucas07}, Pt\cite{Lucas07}
and Au\cite{Portal01, Lucas07}, is energetically more
favorable than the linear chain structure. Thus, we also study the structural, electronic and
magnetic properties of all 4$d$ and 5$d$ transition metal zigzag chains in order to understand
how the physical properties of the monoatomic chains evolve as their structures change from the
linear to zigzag chain.

Relativistic electron spin-orbit coupling (SOC) is the fundamental cause of the orbital
magnetization and also the magnetocrystalline anisotropy energy (MAE) of solids.
The MAE of a magnetic solid is the difference in total electronic energy between
two magnetization directions, or the energy required to rotate the magnetization
from one direction to another. It determines whether a magnet is a hard or soft one.
Furthermore, it acts to reduce the magnitude of superparamagnetic fluctuation in
nanostructures, and hence is a key factor that would determine
whether the nanowires have potential applications in, e.g., high-density recording and
magnetic memory devices. {\it Ab initio} calculations of the MAE have been performed
for mainly the Fe and Co linear chains\cite{Jisang03,Mokrousov05,Jisang06,Aut06}, while
semiempirical tight-binding calculations have been reported for both linear chains and
two-leg ladders of Fe and Co\cite{Druzinic97,Dorantes-Davila98,Aut06}. Very recently,
we have carried out systematic {\it ab initio} calculations of both the MAE and
also the magnetic dipolar (shape) anisotropy energy
for all 3$d$ transition metals in both the linear and zigzag structures.\cite{Tung07}
Remarkably, although the SOC is rather weak in 3$d$ transition metals, compared with 4$d$ and 5$d$ transition
metals, we found that the FM Ni linear chain has a gigantic MAE of $\sim$12 meV/atom.\cite{Tung07}
Therefore, as a continuing endeavor to find nanowires with a large MAE, we have calculated the MAE and the
magnetic dipolar (shape) anisotropy energy for all 4$d$ and 5$d$ transition metals
in both the linear and zigzag structures. Although in this paper we study only free-standing
4$d$ and 5$d$ transition metal chains, the underlying physical trends found may also
hold for monoatomic nanowires created transiently in break junctions\cite{Rodrigues03} or
encapsulated inside 1D nanotubes\cite{Suh03,Mokrousov05} or deposited on weakly interacting
substrates~\cite{Hei04}, {\it albeit}, with the actual values of the physical quantities
being modified.

The rest of this paper is organized as follows. In the next section, we briefly  describe
the theory and computational details we used.
The calculated structural and magnetic properties as well as band structures
of the linear 4$d$ and 5$d$ transition metal chains are presented in Sec. III.
The calculated structural, magnetic and electronic properties of
the zigzag 4$d$ and 5$d$ transition metal chains in both equilibrium and local energy minimum states 
are reported in Sec. IV. The relative stability of the linear and zigzag chain structures is analyzed
in Sec. V. The calculated magnetic anisotropy energies and moments of both
linear and zigzag chains are presented, and also discussed in terms of the calculated
$d$-orbital-decomposed DOSs in Sec. VI. Finally, a summary is given in Sec. VII.

\section{Theory and Computational Method}

In the present calculations, we use the accurate frozen-core
full-potential projector augmented-wave (PAW) method,~\cite{blo94} as implemented
in the Vienna {\it ab initio} simulation package (VASP) \cite{vasp1,vasp2}. The
calculations are based on density functional theory with the
generalized gradient approximation (GGA)\cite{PW91}.
The free-standing atomic chains are modelled by a two-dimensional array of infinite
long, straight or zigzag wires. For both linear and zigzag chains, the nearest wire-wire
distance between the neighboring chains is, at least, 15 \AA, which should be wide enough
to decouple the neighboring wires. A large plane-wave cutoff energy of $\sim$350 eV is
used for all 4$d$ and 5$d$ transition metal chains.

The equilibrium bond length (lattice constant) of the linear atomic chains in the nonmagnetic
(NM), ferromagnetic (FM) and antiferromagnetic (AF) states is
determined by locating the minimum in the calculated total energy as a function of the
interatomic distance. The results are also compared with that obtained by structural
optimizations, and the differences are small (within 0.4 \%) for, e.g., the
Ru, Rh and Pd chains. For the zigzag chains,
the theoretical atomic structure is determined by structural relaxations using the conjugate
gradient method. The equilibrium structure is obtained when all the forces acting on the atoms
and the axial stress are less than 0.02 eV/\AA$ $ and 2.0 kBar, respectively.
The $\Gamma$-centered Monkhorst-Pack scheme with a $k$-mesh
of $1\times1\times n$ ($ n = 40$) in the full Brillouin zone (BZ),
in conjunction with the Fermi-Dirac-smearing method with $\sigma = 0.01$ eV,
is used to generate $k$-points for the BZ integration.
With this $k$-point mesh, the total energy is found to converge to within 10$^{-3}$ eV.

Because of its smallness, {\it ab initio} calculation of the MAE
is computationally very demanding and needs to be carefully carried out (see, e.g.,
Refs. \onlinecite{Daalderop2, guo91}). A very fine $k$-point mesh with $n$ being 200
for both the linear and zigzag chains, is used. The same $k$-point mesh is used for
the band structure and density of states calculations.
As in our previous publication on the 3$d$ TM chains\cite{Tung07},
we use the force theorem approach to calculate the MAE,
i.e., the MAE is calculated as the total energy difference between the two
relativistic band structure calculations for the two different magnetization
directions (e.g., parallel and perpendicular to the chain) concerned
using the frozen charge density obtained in a prior self-consistent
scalar relativistic calculation.\cite{vasp} The total energy
convergence criteria is 10$^{-7}$ eV/atom.

\begin{figure}
\includegraphics[width=6cm]{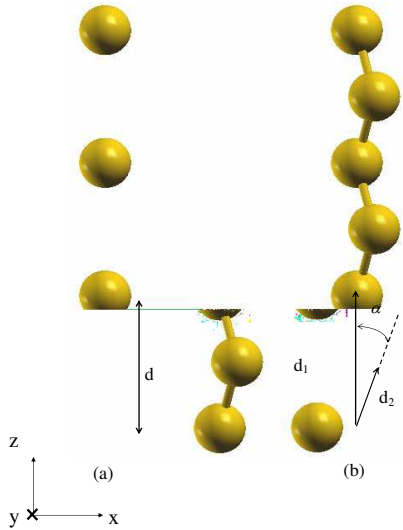}\\
\caption{(color online) Schematic structure diagram for (a) the linear and (b) zigzag atomic chains.}
\end{figure}

\section{Linear Chains}

\begin{figure}
\includegraphics[width=7cm]{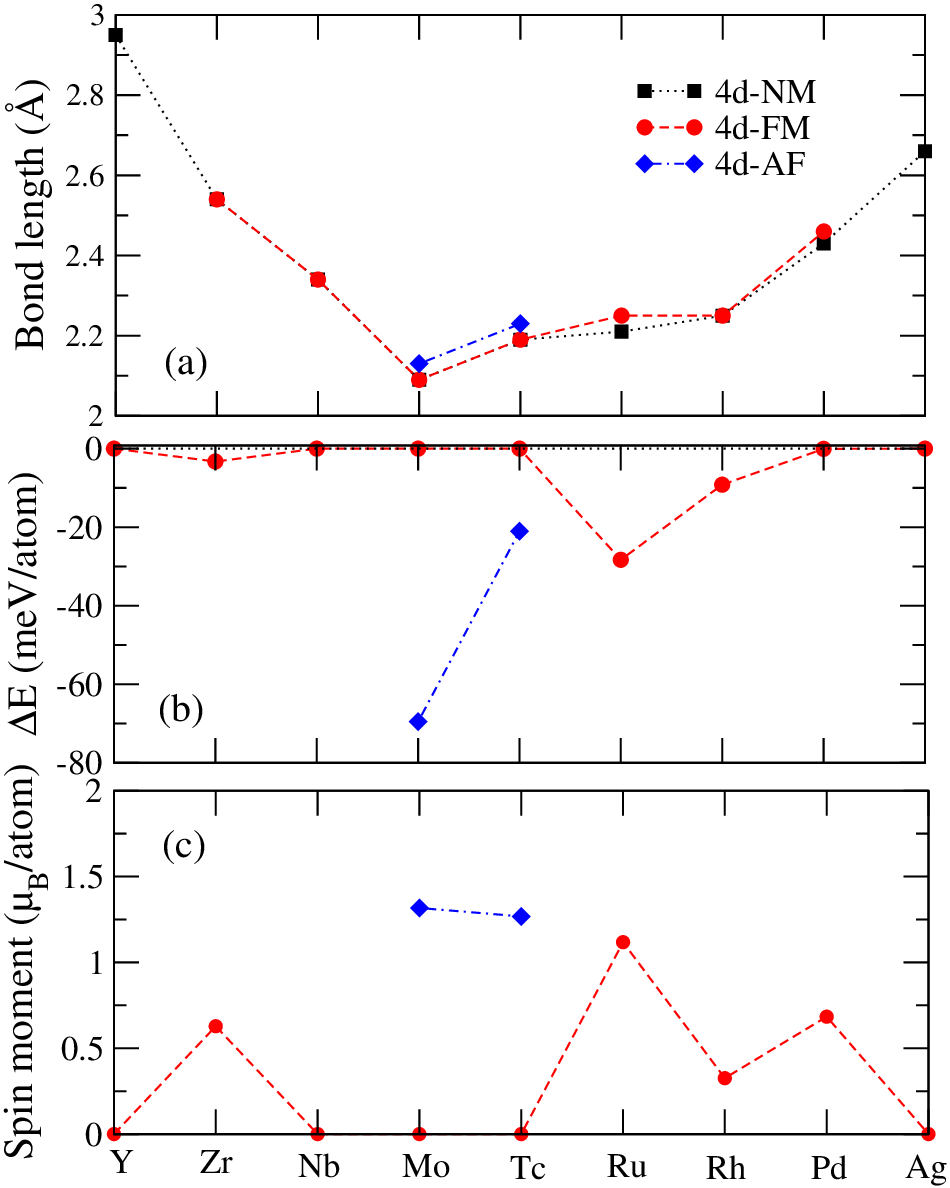} \\
\caption{(color online) (a) Equilibrium bond length ({\AA}), (b) magnetization energy
($\Delta E$) (i.e., the total energy of a magnetic state relative to that of nonmagnetic state)
($\Delta E = E^{FM(AF)} - E^{NM}$) and (c) spin magnetic moments ($\mu_B$) of all the 4$d$
TM linear atomic chains in the NM, FM, and AF states.}
\end{figure}

\begin{figure}
\includegraphics[width=7cm]{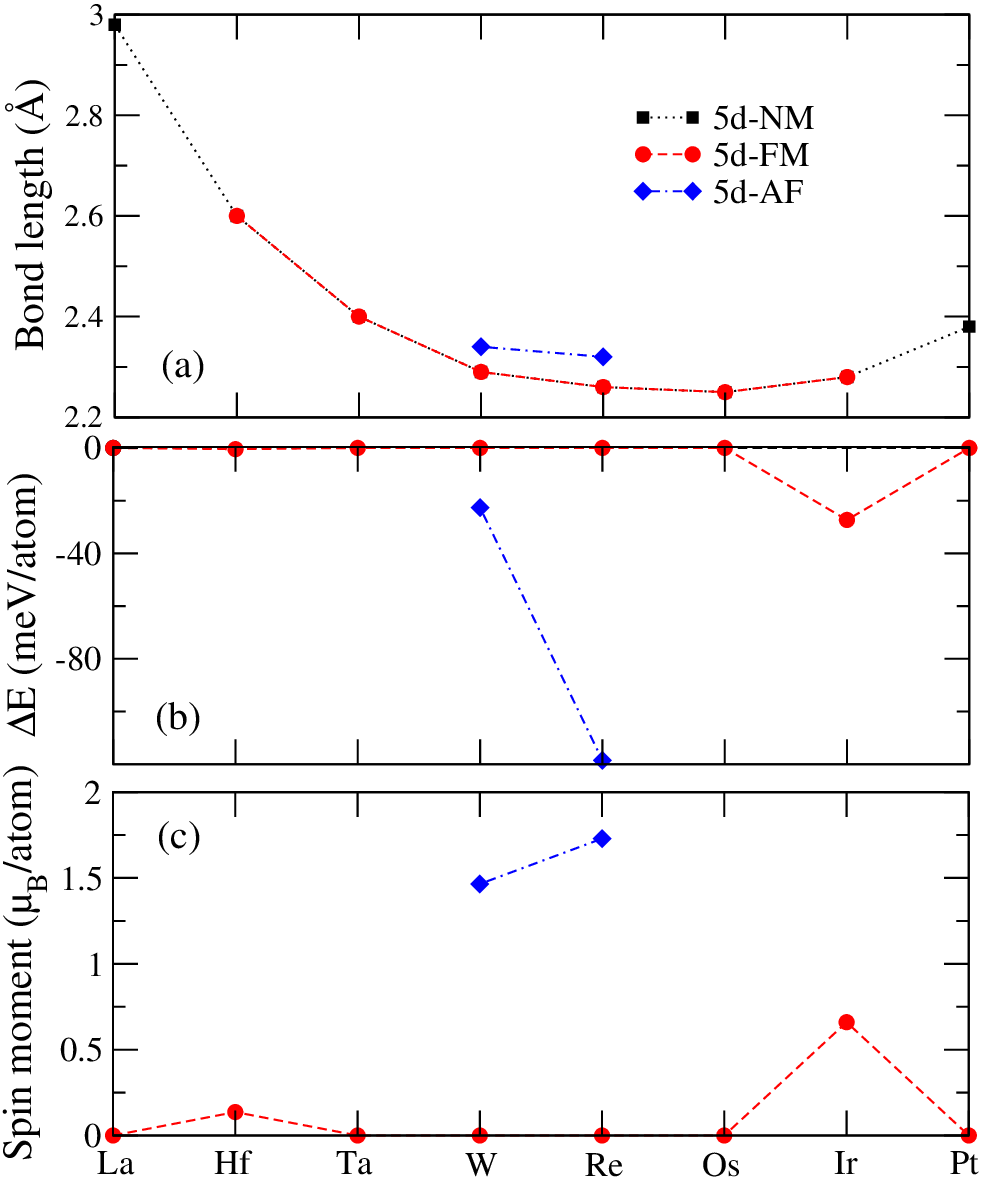} \\
\caption{(color online) (a) Equilibrium bond length ({\AA}), (b) magnetization energy
($\Delta E$) (i.e., the total energy of a magnetic state relative to that of nonmagnetic state)
($\Delta E = E^{FM(AF)} - E^{NM}$) and (c) spin magnetic moments ($\mu_B$) of all the 5$d$
TM linear atomic chains in the NM, FM, and AF states.}
\end{figure}

\subsection{Magnetic state and spin magnetic moment}

The calculated equilibrium bond lengths ($d$) and spin magnetic moments
of all the 4$d$ and 5$d$ transition metal linear chains in the NM, FM and
AF states are displayed in Fig. 2, and Fig. 3, respectively. They are also
listed in Table I. The calculated total energy relative to that of the NM
state (i.e., the magnetization energy) of the FM and
AF linear atomic chains are also shown in Fig. 2, Fig. 3 and Table I.
It is clear from Figs. 2 and 3 that all of the 4$d$ and 5$d$ TM elements
except Y, Nb, La, Ta, Os and Pt, become magnetic in the linear chain structure.
Furthermore, for all the 4$d$ and 5$d$ TM elements, except Y, Nb, La, Ta, Os and Pt,
NM state is unstable and the ground state is either FM and AF (see Fig. 2,
Fig. 3 and Table I). Among the 4{\it d} TM linear chains, the ground state for
the Zr, Ru, Rh, and Pd chains is ferromagnetic while that for the Mo, and
Tc chains is antiferromagnetic. For the 5$d$ TM linear chains, the ground state
for the Hf and Ir chains is ferromagnetic and the ground state for the Re and W
are antiferromagnetic. The Y, Nb, La, Ta, Os and Pt chains are nonmagnetic at the
equilibrium bond length. We recently reported\cite{Tung07} that in the 3{\it d} TM
linear chains, the equilibrium bond length in a magnetic state is significantly
larger than that in the nonmagnetic state. For example, the magnetization induced
increase in the bond length in the Cr chain is 54 \%. In contrast,
Fig. 2, Fig. 3 and Table I show that in the 4$d$ and 5$d$ TM linear chains,
the difference in bond length between a magnetic (FM or AF) state and the NM state
is much smaller. The largest lattice magnetolattice expansion occurs in the AF Re
chain but it amounts only to 3 \%. This is due to much weak magnetization in the
4$d$ and 5$d$ linear chains, as indicated by the smaller magnetic moments and much
smaller magnetization energies in these atomic chains at equilibrium (Table I).

To see how the magnetic properties of the atomic chains evolve with the interatomic
distance, we plot the spin and orbital moments for some 4$d$ (Y, Zr, Nb, and Pd)
and 5$d$ (La, Hf, Ta, Os, Ir and Pt) TM chains in the FM state as a function of
the bond length in Fig. 4. For most selected ferromagnetic TM (except Y, La, Zr and Hf) chains,
the spin moment generally becomes larger as the bond length is increased from
its equilibrium value (Table I).Interestingly, the spin moment of the Hf chain, in contrast, decreases monotonically
when the chain is elongated, and eventually disappears at the bond length of
2.7 \AA$ $ [Fig. 4(c)]. The spin moment of the Zr chain decreases slightly as the bond
length increases, but increases again when the bond length goes beyond $\sim$2.7 \AA.
Surprisingly, when the Y (La) chain is sufficiently compressed
[at the bond length of $\sim$2.35 (2.65) \AA],
the ferromagnetism appears, and the spin moment increases as the chain is further
compressed. Finally, for the Y, La, Ta, Os and Pt chains, the sufficient elongation
of the bond length would induce a FM state (Fig. 4).

Our calculated bond lengths, spin moments and magnetization energies generally agree rather
well with available previous {\it ab initio}
calculations~\cite{Mokrousov05,Mokrousov06,Nautiyal04,Delin03,Delin04,Delin06,Spisak03}.
Nonetheless, a few notable differences exist. For example,
our calculated bond length (2.13 \AA) of the AF Mo chain is 7.0 \% smaller than that (2.28 \AA)
reported in Ref. \onlinecite{Mokrousov06} but in good agreement with
Ref. \onlinecite{Spisak03} (2.15 \AA). Also, our equilibrium bond
lengths of the Os and Pt chains are smaller than that from Ref~\onlinecite{Delin03},
but the differences are within 2.7 \%. Another notable difference is that our calculations
suggest that the Os chain is nonmagnetic in equilibrium but become ferromagnetic
only when the bond length is larger than $\sim$2.55\AA$ $ (Fig. 4), while, according to
Ref. \onlinecite{Delin03}, it is ferromagnetic at the equilibrium bond length.
Our calculated magnetization energies (Table I) for the AF Mo and Tc chains
are smaller than that reported in Ref. \onlinecite{Mokrousov06}
(197 and 53 meV/atom, respectively) and in Ref. \onlinecite{Spisak03}
(92 and 65 meV/atom, respectively).

\begin{table}
\caption{Equilibrium bond lengths ($d$) (in \AA), total energies ($E_t$)
(in meV/atom) in the FM and AF states (relative to the NM state), and spin magnetic
moments ($m_s$) (in $\mu_B$/atom), of the 4$d$ and 5$d$ transition metal linear chains
from scalar relativistic calculations.}
\begin{ruledtabular}
\begin{tabular}{cccccccc}
   &$d_{NM}$ &$E_t^{FM}$ &$m_s^{FM}$ &$d_{FM}$ &$E_t^{AF}$ &$m_s^{AF}$ & $d_{AF}$ \\ \hline
   &         &           &           &  4{\it d} metals &&& \\
Y  & 2.95 &          &         &       &           &       &      \\
Zr & 2.54 &  -3.29 &  0.628  & 2.54  &           &       &      \\
Nb & 2.34 &        &         &       &           &       &      \\
Mo & 2.09 &        &         &       &  -69.53  & 1.317 & 2.13 \\
Tc & 2.19 &        &         &       &  -21.02  & 1.268 & 2.23 \\
Ru & 2.21 & -28.30 &  1.118  & 2.25  &           &       &      \\
Rh & 2.25 &  -9.19 &  0.328  & 2.25  &           &       &      \\
Pd & 2.43 &  -0.05 &  0.684  & 2.46  &           &       &      \\
Ag & 2.66 &        &         &       &           &       &      \\
   &         &           &           &  5{\it d} metals &&& \\
La & 2.98 &        &         &       &           &       &      \\
Hf & 2.60 &  -0.46 &  0.137  & 2.60  &           &       &      \\
Ta & 2.40 &        &         &       &           &       &      \\
W  & 2.29 &        &         &       &   -22.59  & 1.465 & 2.34 \\
Re & 2.26 &        &         &       &  -118.52  & 1.729 & 2.32 \\
Os & 2.25 &        &         &       &           &       &      \\ 
Ir & 2.28 & -27.32 &  0.660  & 2.28  &           &       &      \\
Pt & 2.38 &        &         &       &           &       &      \\
Au & 2.60 &        &         &       &           &       &      \\

\end{tabular}
\end{ruledtabular}
\end{table}

\begin{table}
\caption{Spin ($m_s$) and orbital ($m_o$) magnetic moments (in $\mu_B$/atom) of the magnetic
4$d$ and 5$d$ transition metal linear chains at the equilibrium bond lengths (Table I)
with magnetization parallel ({\bf m}$\parallel \hat{z}$) and perpendicular ({\bf m}$\perp \hat{z}$)
to the chain axis from fully relativistic charge selfconsistent calculations.}
\begin{ruledtabular}
\begin{tabular}{cccccc}
   &      &\multicolumn{2}{c}{{\bf m}$\parallel \hat{z}$} &\multicolumn{2}{c}{{\bf m}$\perp \hat{z}$}\\
   &      & $m_s$  & $m_o$   & $m_s$ & $m_o$  \\ \hline
          \multicolumn{6}{c}{4{\it d} metals} \\
Zr & (FM) & 0.631  & -0.065  & 0.610 & -0.007 \\
Mo & (AF) & 1.337  & -0.008  & 1.181 &  0.005 \\
Tc & (AF) & 1.353  &  0.463  & 1.252 &  0.046 \\
Ru & (FM) & 1.115  & -0.106  & 1.076 &  0.058 \\
Rh & (FM) & 0.317  &  0.428  & 0.017 &  0.000 \\
Pd & (FM) & 0.345  & -0.043  & 0.636 &  0.126 \\
          \multicolumn{6}{c}{5{\it d} metals} \\
Hf & (FM) & 0.235  & -0.198  &       &        \\
W  & (AF) & 1.184  & -0.307  & 1.371 & -0.005 \\
Re & (AF) & 1.564  &  0.115  & 1.644 &  0.146 \\
Os & (FM) &        &         & 0.444 &  0.046 \\
Pt & 2.38 & 0.124  & 0.100   &       &        \\
\end{tabular}
\end{ruledtabular}
\end{table}

\begin{figure}
\includegraphics[width=8cm]{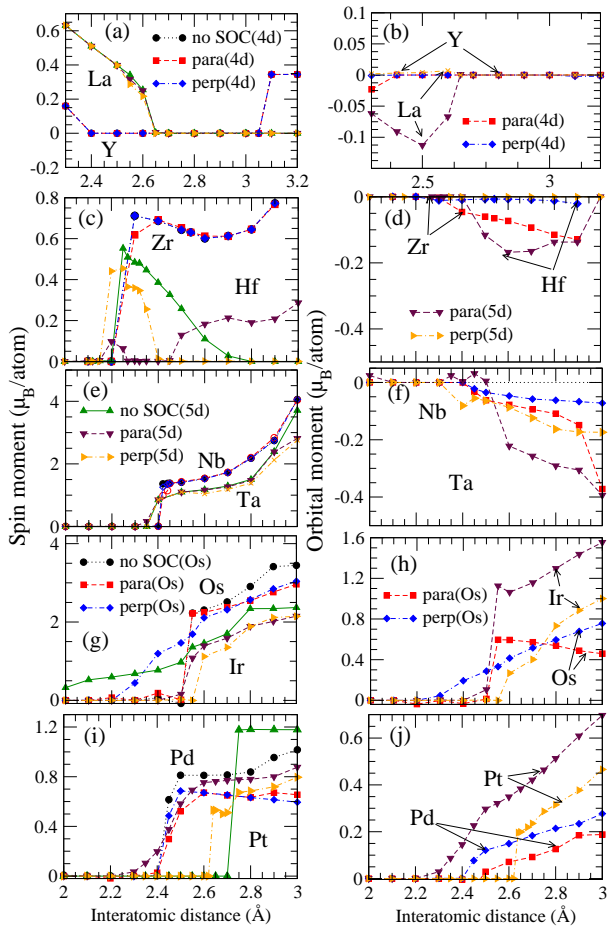} \\
\caption{(color online) Spin (left panels) and orbital (right panels) magnetic moments as a
function of interatomic distance of the ferromagnetic Y, La, Zr, Hf, Nb, Ta, Os, Ir
Pd and Pt linear chains. In the left panels, "no-SOC" means the results from the scalar
relativistic calculations. "Para" ("perp") denotes the magnetization being parallel (perpendicular)
to the chain axis. The spin magnetic moment for the Pd chain goes to zero at $\sim$ 3.6 \AA.}
\end{figure}

\subsection{Orbital magnetic moment and colossal magnetic anisotropy}

The spin and orbital magnetic moments in the magnetic 4$d$ and 5$d$ TM atomic chains
in equilibrium from the fully relativistic charge selfconsistent calculations are listed in Table II.
We note that the SOC affect slightly the spin moments in the AF TM chains
and also FM Zr and Ru chains (see Tables I and II).
However, in the other cases, the spin magnetic moments in Table II are generally much smaller than that
obtained from the scalar-relativistic calculations (Table I),
unlike in the 3$d$ TM chains where the SOC hardly affects the spin magnetic moments\cite{Tung07},
In fact, the SOC completely suppresses the spin magnetic moment
in the FM Ir chain in equilibrium (Tables I and II).
Interestingly, the SOC-induced reduction of the spin magnetic moment is magnetization-direction
dependent. Table II shows that the spin moment of the Rh chain with magnetization
parallel to the chain axis remains almost unchanged while that perpendicular to the
axis becomes nearly diminished. In the Pd chain, in contrast, the spin moment for
magnetization along the axis decreases nearly by half while that perpendicular to the
axis remains nearly unchanged (Table II). Dramatically, in the FM Hf chain, the SOC
fully suppresses the magnetization when the magnetization is perpendicular to the
chain axis, but it nearly doubles the spin moment when the magnetization is along
the axis. This interesting magnetic anisotropy is called the colossal magnetic anisotropy
(CMA) by Smogunov {\it et al.}\cite{smo08}, who reported recently this
CMA in the Pt chain.
The CMA means that a magnetization magnitude  could be finite only along certain
directions and also that it is strictly impossible to rotate magnetization
into certain directions. Earlier calculations\cite{Mokrousov06} also suggested the CMA to
occur in the Rh linear chain.
Our calculations here not only corroborate this finding of Smogunov {\it et al.}\cite{smo08}
but also reveal the CMA in other 5$d$ transition metal linear chains such as Hf and Os
(Table II).

When the SOC is not taken into account, the spin moment in the 4$d$ and 5$d$ TM linear chains
generally increases monotonically as the bond length is increased, as shown in Fig. 4.
However, the behavior of the magnetic properties of the 5$d$ TM linear chains under the influence
of the SOC is very different from that of the 3$d$ TM chains. For example, from the scalar relativistic
calculations, the Ir chain at the interatomic distance starting from 2.0 to 3.0 \AA,
has a finite magnetic moment in the range of 0.4$\sim$2.4 $\mu_B$/atom (Fig. 4g).
When the SOC is taken into account, the Ir chain becomes nonmagnetic when the interatomic distance is smaller
than 2.5 \AA, but has a finite magnetic moment when the interatomic distance larger then 2.5 \AA.
Similar behavior can also be seen in the Os, and Pt chains (Fig. 4 g and i).
Our scalar relativistic calculations show that the Os chain is nonmagnetic if the interatomic distance is below
2.4 \AA, whilst our fully relativistic calculations indicate that, for the magnetization perpendicular
to the chain direction, it become magnetic at the interatomic distance above 2.2 \AA.
In contrast, for the axial magnetization, the Os chain would become ferromagnetic only
when the interatomic distance is larger than 2.5 \AA. Therefore, the Os chain exhibits the
CMA~\cite{smo08} when the interatomic distance falls between 2.2 and 2.5 \AA.
Fig. 4 further shows that the Hf, Ir and Pt linear chains also exhibit the CMA in the interatomic distance of
2.25$\sim$3.0 \AA$ $(Hf), 2.52$\sim$2.78\AA$ $(Ir) and 2.30$\sim$2.63 \AA$ $(Pt), respectively.

The SOC provides the essential symmetry breaking that gives rise to orbital magnetization
in magnetic solids.
When the SOC is included in our calculations, the calculated orbital magnetic moments
in the FM 4$d$ TM chains at equilibrium bond length
are listed in Table II.
Surprisingly, even for the 5$d$ TM linear atomic chains, the calculated orbital magnetic moments
are not large.
For example, the calculated orbital moments in the 5$d$ TM linear atomic chains are within
$\sim$ 0.2 $\mu_B$/atom (Table II). The calculated orbital moments in the 4$d$ TM linear chains
in equilibrium can be larger, e.g., being $\sim$ 0.2 $\mu_B$/atom in the AF Tc and FM Rh chains
with the magnetization along the chain direction (Table II).
respectively. Therefore, although the SOC is stronger in 4$d$ and 5$d$ transition metals
than in 3$d$ ones, the calculated orbital magnetic moments in the 4$d$ and 5$d$ transition metals
chains at the equilibrium bondlength is not necessarily larger than in the 3$d$ transition
metal chains.\cite{Tung07}
As for the spin moments, the magnitude of the orbital moments generally increases monotonically
with the bond length, as can
be seen in Fig. 4, with one notable exception of the La chain (Fig. 4b).
The orbital moment shows a strong dependence on the magnetization
orientation (Fig. 4, right panels). As in the 3$d$ TM chains\cite{Tung07}, the orbital moment in
the 4$d$ and 5$d$ TM chains with the magnetization along the chain direction
is usually much higher than that for the magnetization perpendicular to the chain.
However, in the Pd chain, the orbital moment with the magnetization along the chain direction
is significantly larger than that for the magnetization perpendicular to the chain. (Fig. 4j).

\subsection{Band structures and density of states}
\begin{figure}
\includegraphics[width=8cm]{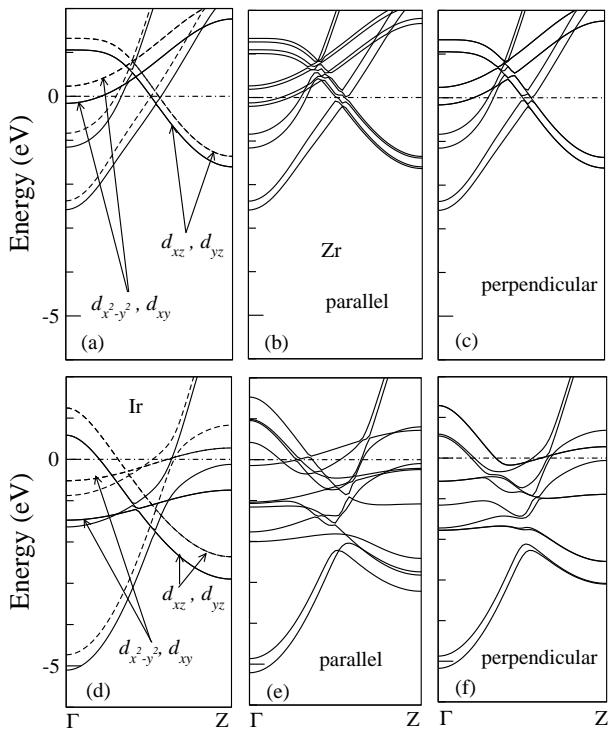}
\caption{Band structures of the Zr (upper panels) and Ir (lower panels) linear chains at
2.6 \AA. Left panels: the scalar-relativistic band structures; the middle and right panels: the fully
relativistic band structures with the the magnetization parallel to and perpendicular to the chain axis,
respectively. In the left panels, the solid and dashed lines represent (spin up)
and (spin down) bands, respectively. The Fermi level (the dotted horizontal line)
is at the zero energy.}
\end{figure}

Let us now examine the band structure of selected transition metal linear chains
in order to understand the calculated magnetic properties.
The energy bands obtained without and also with the SOC for
the Zr and Ir chains in the FM state at 2.6 \AA$ $ are plotted in Fig. 5.
In the absence of the SOC, because of the uniaxial rotational symmetry, the bands
may be grouped into three sets, namely, the nondegenerate $s$- and $d_{z^2}$-dominant
bands, double degenerate ({\it d$_{ xz}$, d$_{yz}$}), and ({\it d$_{x^2-y^2}$, d$_{xy}$})
dominant bands (see the left panels in Fig. 5). The ({\it d$_{x^2-y^2}$, d$_{xy}$})
bands are narrow because the $d_{x^2-y^2}$ and $d_{xy}$ orbitals are perpendicular to the
chain, thus forming weak $\delta$ bonds. The ({\it d$_{xz}$, d$_{yz}$}) bands, on the other
hands, are more dispersive due to the stronger overlap of the $d_{xz}$ and  $d_{yz}$ orbitals
along the chain, which gives rise to the $\pi$ bonds. The $s$- and $d_{z^2}$ dominant
bands are most dispersive since these orbitals form strong $\sigma$ bonds along the chain.
The left panels in Fig. 5 show that the less dispersive {\it d$_{x^2-y^2}$, d$_{xy}$}
bands are near the Fermi level and spin-split. In the Zr linear chain, one spin-split
{\it d$_{x^2-y^2}$, d$_{xy}$} band is partially occupied near the $\Gamma$-point while the
other band is completely empty. In the Ir chain, one split band lies completely below the
Fermi level while the other band is only partially occupied. Thus, the relatively narrow
{\it d$_{x^2-y^2}$, d$_{xy}$} bands play an important role in magnetism, and that is the
main reason why Zr and Ir chains are ferromagnetic at the bondlength of 2.6 \AA.

 The directional dependence of the orbital magnetization can be explained by analyzing
the fully relativistic band structures (see Fig. 5). For the Zr linear chain with
the axial magnetization (Fig. 5b), the doubly degenerate {\it d$_{x^2-y^2}$, d$_{xy}$} bands
are split into two with angular momenta {\it $m_l$= $\pm$}2.
If one of them is fully occupied and the other is empty, the resulting orbital moment is 2.
Nonetheless, in the Zr linear chain, both are partially occupied with different occupation numbers
(Fig. 5b), resulting in an orbital moment of -0.07 $\mu_B$/atom. Of course,
the larger the SOC splitting, the larger the difference in the occupation number and
hence the larger the orbital moment. Therefore, the Ir chain has a larger axial orbital
moment (1.06 $\mu_B$/atom), because one of the split {\it d$_{x^2-y^2}$, d$_{xy}$} bands
lies almost completely below the Fermi level (see Fig. 5e).
However, for the perpendicular magnetization, the {\it d$_{x^2-y^2}$, d$_{xy}$} bands remain degenerate
(Fig. 5c and Fig. 5f) and hence do not contribute to the orbital magnetization.
Nonetheless, as pointed out in Ref. \onlinecite{Mokrousov06},
the SO-split {\it d$_{x^2-y^2}$, d$_{xy}$}-($d_{xz}$, $d_{yz}$) bands near the Fermi level
would hybridize (Fig. 5f) and this hybridization would give rise to a smaller perpendicular orbital moment
of 0.27  $\mu_B$/atom in the Ir linear chain. For the Zr chain, this
hybridization does not occur near the Fermi energy (Fig. 5c).
Therefore, the Zr chain have a tiny orbital moment of -0.01 $\mu_B$/atom
when the magnetization is perpendicular to the chain axis.
Of course, when the SOC is included, the degenerate {\it d$_{xz}$, d$_{yz}$} bands are also split into
the $m_l= -1$ and +1 bands for the axial magnetization, but remain degenerate for the perpendicular
magnetization (see Fig. 5). This SOC splitting of the ($d_{xz}$, $d_{yz}$) band and ($d_{x^2-y^2}$, $d_{xy}$) band is
proportional to $|<d_{xz}|H_{SO}|d_{yz}>|^2$ and $|<d_{x^2-y^2}|H_{SO}|d_{xy}>|^2$,
respectively. Here $H_{SO}$ is the SOC Hamiltonian. Since
 $|<d_{xz}|H_{SO}|d_{yz}>|^2$:$|<d_{x^2-y^2}|H_{SO}|d_{xy}>|^2$ = 1:4,~\cite{tak76}
the SOC splitting of the ($d_{xz}$, $d_{yz}$) bands is much smaller than
the ($d_{x^2-y^2}$, $d_{xy}$) bands (see Fig. 5). Therefore, the ($d_{xz}$, $d_{yz}$) bands
would make a much smaller contribution to the orbital magnetization

Electric and spin current transports are determined by the characteristics of the
band structure near the Fermi level ($E_F$) in the systems concerned. Therefore, it would
be interesting to examine the energy bands and density of states (DOS) of the atomic
chains in the vicinity of the $E_F$. The spin-decomposed DOS for all the 4$d$ and 5$d$
linear chains in equilibrium are displayed in Fig. 7 and Fig. 8, respectively.
For the FM Zr, Ru, Rh, Pd, Hf and Ir chains, the density of states at the $E_F$ are
spin-polarized (Fig. 7 and Fig. 8).
This is usually quantified by the spin-polarization $P$ defined as
\begin{equation}
P=\frac{N_{\uparrow}(E_F)-N_{\downarrow}(E_F)}{N_{\uparrow}(E_F)+N_{\downarrow}(E_F)},
\end{equation}
where $N_{\uparrow}(E_F)$ and $N_{\downarrow}(E_F)$ are the spin-up and spin-down DOS
at the $E_F$, respectively.
The most useful materials for the spintronic applications are the so-called half-metallic
materials in which one spin channel is metallic and the other spin channel is insulating.
The spin-polarization for these half-metals is either 1.0 or -1.0, and the electric conduction
would be fully spin-polarized. The calculated spin-polarization and also the numbers of the conduction
bands that cross the Fermi level in the 4$d$ and 5$d$ TM chains are listed in Table III.
It is clear that the $P$ of the FM Zr, Ru, Rh, and Ir linear chains is rather large ($\ge$ 0.4), though still
smaller than many 3$d$ TM linear chains.\cite{Tung07} None of the 4$d$ and 5$d$ TM linear chains
in the FM state is half-metallic.
Interestingly, the FM Zr chain has a positive spin polarization, while the Ru, Rh, Hf, Pd and Ir chains
have a negative spin polarization (Table II, Fig. 7 and Fig. 8).

\begin{table}
\caption{Numbers ($n_c^\uparrow$ and $n_c^\downarrow$) of the spin-up and spin-down
conduction bands crossing the Fermi level, and spin-polarization $P$ at the Fermi level
for the 4$d$ and 5$d$ TM atomic chains in the FM state. }
\begin{ruledtabular}
\begin{tabular}{ccccc}
 & \multicolumn{2}{c}{linear chain}&\multicolumn{2}{c}{zigzag chain } \\
  &($n^{\uparrow}_{c}$, $n^{\downarrow}_{c}$) & $P$ &($n^{\uparrow}_{c}$, $n^{\downarrow}_{c}$) & $P$ \\ \hline
      &        &      4$d$ metals    &     &               \\
Y     &        &                     &(2,3)&      0.09     \\
Zr    & (4,3)  &  0.48               &(3,4)&     -0.09     \\
Mo    &        &                     &(6,8)&     -0.11     \\
Ru    & (3,4)  & -0.41              &     &               \\
Rh    & (3,4)  & -0.65               &(3,6)&     -0.53     \\
Pd    & (3,4)  & -0.22               &(3,6)&     -0.30     \\ \hline
      &        &      5$d$ metals    &     &               \\
Hf    & (4,3)  & -0.31              &     &               \\
W     &        &                     &(6,8)&     -0.16    \\
Re    &        &                     &(2,4)&     -0.38    \\
Os    &        &                     &(5,7)&     -0.10     \\
Ir    & (3,4)  &  -0.47              &(7,9)&     -0.28    \\

\end{tabular}
\end{ruledtabular}
\end{table}

\begin{figure}
\includegraphics[width=8cm]{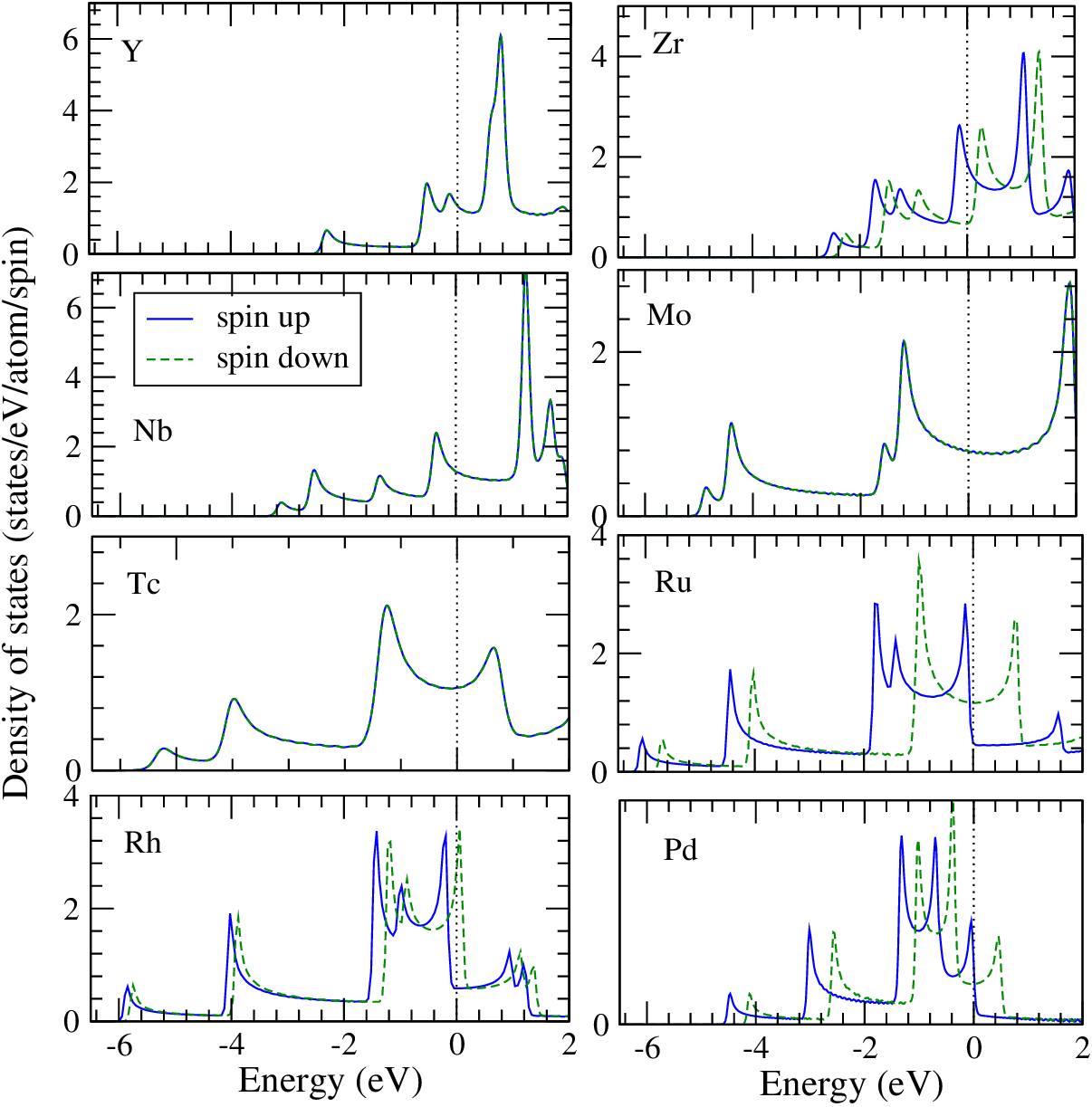}
\caption{(color online) Density of states of the FM 4{\it d} TM linear atomic chains at the equilibrium
bond length. The Fermi level (dotted vertical lines) is at the zero energy.}
\end{figure}

\begin{figure}
\includegraphics[width=8cm]{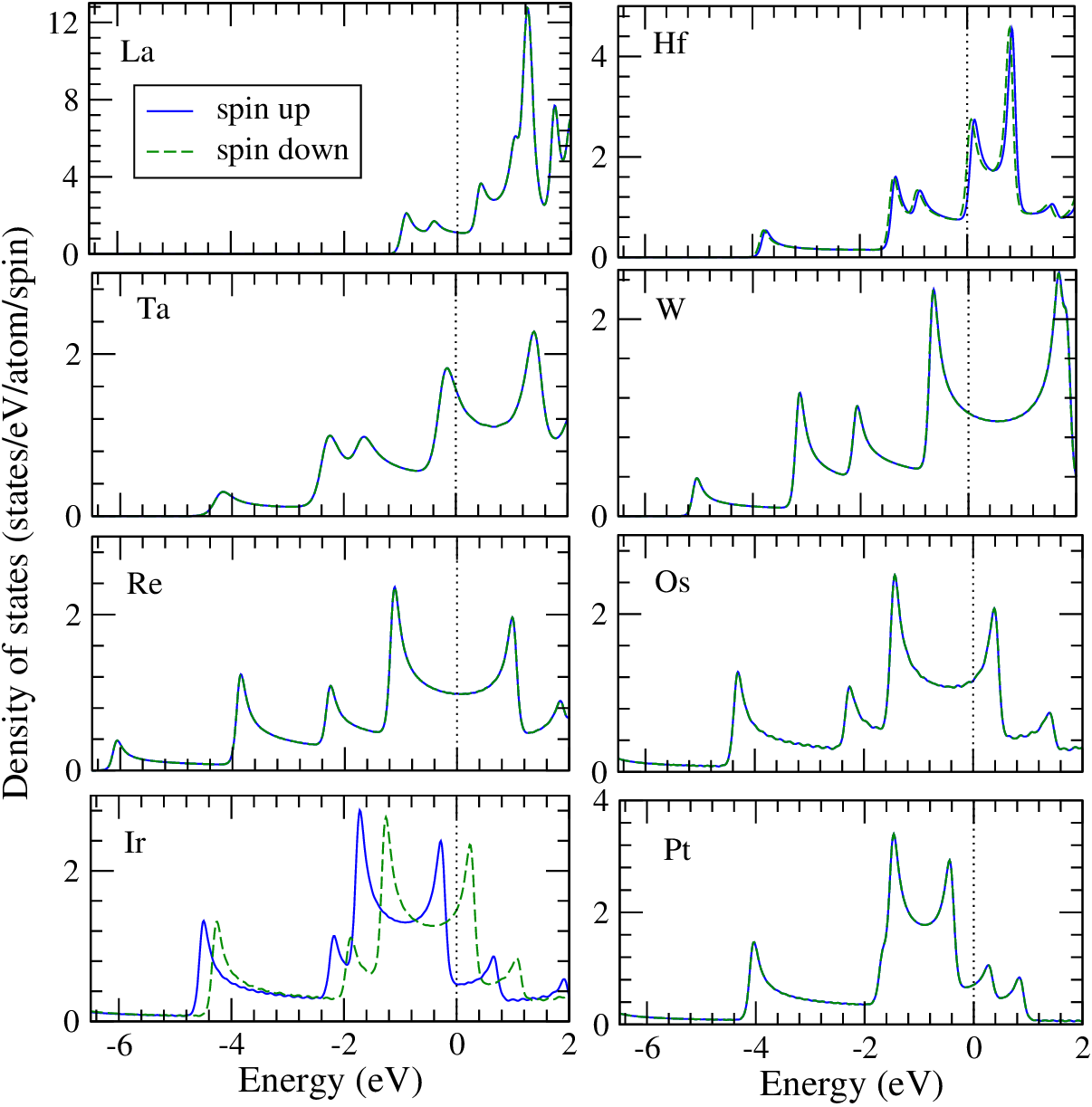}
\caption{(color online) Density of states of the FM 5{\it d} TM linear atomic chains at the equilibrium
bond length. The Fermi level (dotted vertical lines) is at the zero energy.}
\end{figure}

\section{Zigzag Chains}

The zigzag structure for metal monoatomic wires has already been observed in experiments~\cite{whitman91}.
Among 4$d$ and 5$d$ transition metals, structural~\cite{Ribeiro03,lin04} and
magnetic~\cite{Lucas07} properties of Zr, Rh, Pd, W, Ir, and Pt zigzag atomic chains have been
studied theoretically in recent years. In the present paper, we perform a systematic
{\it ab initio} study of the structural, electronic and magnetic properties of the zigzag chain
structure of all the 4$d$ and 5$d$ transition metals.

\subsection{Structure and magnetic moments}

\begin{table}
\caption{Equilibrium structural parameters (see Fig. 1b for symbols $d_1$, $d_2$, $\alpha$),
spin magnetic moment ($m_s$) and magnetization energy ($\Delta E$)
of the 4$d$ and 5$d$ transition metal zigzag chains from  the scalar relativistic calculations.
$d_1$ and $d_2$ are in the unit of \AA, and $\alpha$ is in the unit of degree.
$\Delta E$ is in the unit of meV/atom, and $m_s$ in the unit of $\mu_B$/atom.}
\begin{ruledtabular}
\begin{tabular}{ccccccc}
     &      & $d_1$ & $d_2$ &$\alpha$& $m_s$ &$\Delta E$  \\ \hline
     &      &       &   4$d$ metals  &       &       &        \\
 Y   & (NM) & 3.12  & 3.03  & 59.0   &       &            \\
     & (FM) & 3.17  & 3.08  & 59.0   & 0.482 & 388.1      \\
Zr   & (NM) & 2.81  & 2.71  & 58.7   &       &            \\
     & (FM) & 2.85  & 2.74  & 58.6   & 0.162 & -2.93      \\
Nb   & (NM) & 2.51  & 2.63  & 61.5   &       &            \\
Mo   & (NM) & 2.38  & 2.51  & 61.7   &       &            \\
     & (FM) & 2.45  & 2.53  & 61.0   & 0.267 & -2.90      \\
Tc   & (NM) & 2.40  & 2.47  & 60.9   &       &            \\
Ru   & (NM) & 2.40  & 2.47  & 60.9   &       &            \\
     & (FM) & 2.49  & 2.48  & 59.5   & 1.526 & -26.0      \\
     & (AF) & 2.41  & 2.46  & 60.7   & 0.306 & -5.9       \\
Rh   & (NM) & 2.39  & 2.60  & 62.6   &       &            \\
     & (FM) & 2.59  & 2.49  & 59.1   & 1.355 & -30.0      \\
Pd   & (NM) & 2.56  & 2.64  & 61.0   &       &            \\
     & (FM) & 2.55  & 2.66  & 61.3   & 0.392 & -0.7       \\
     & (AF) & 2.56  & 2.64  & 60.9   & 0.266 & -2.2       \\
Ag   & (NM) & 2.73  & 2.78  & 60.9   &       &            \\

     &      &       &   5$d$ metals  &       &       &    \\
La   & (NM) & 3.24  & 3.10  & 58.5   &       &            \\
Hf   & (NM) & 2.89  & 2.71  & 57.7   &       &            \\
Ta   & (NM) & 2.75  & 2.51  & 56.7   &       &            \\
 W   & (NM) & 2.48  & 2.56  & 61.0   &       &            \\
     & (FM) & 2.48  & 2.56  & 61.0   & 0.262 & -1.8       \\
Re   & (NM) & 3.22  & 2.25  & 48.8   &       &            \\
     & (FM) & 3.22  & 2.25  & 44.3   & 0.516 & -259.5     \\
Os   & (NM) & 2.49  & 2.44  & 59.3   &       &            \\
     & (FM) & 2.50  & 2.44  & 59.1   & 0.457 & -26.7      \\
     & (AF) & 2.50  & 2.44  & 59.2   & 0.360 & -7.9       \\
Ir   & (NM) & 2.45  & 2.53  & 61.0   &       &            \\
     & (FM) & 2.44  & 2.56  & 61.5   & 0.629 & -36.7      \\
Pt   & (NM) & 2.49  & 2.65  & 62.0   &       &            \\
Au   & (NM) & 2.67  & 2.76  & 61.1   &       &            \\
\end{tabular}
\end{ruledtabular}
\end{table}

\begin{table}
\caption{Structural parameters ($d_1$, $d_2$, $\alpha$) (see Fig. 1b)
of the zigzag chain at the second local energy minimum state.
$d_1$ and $d_2$ are in the unit of \AA, and $\alpha$ is in the unit of degree.
$\Delta E$ (meV/atom) is the energy difference between the second energy
minimum and the corresponding energy minimum listed in Table IV.
The second local minimum state of the Zr and Ir chains only is ferromagnetic
with a spin moment of 0.295 and 0.285 $\mu_B$/atom, respectively.
The elements whose zigzag chains do not have the second energy 
minimum are not listed here}
\begin{ruledtabular}
\begin{tabular}{cccccc}
     &      & $d_1$ & $d_2$ &$\alpha$ & $\Delta E$ \\ \hline
     &      &       &   4$d$ metals  &    &     \\
Zr   & (FM) &  4.25 & 2.44 &  29.6 & 1015  \\
Nb   & (NM) &  2.80 & 2.45 &  55.1 & -41.6  \\
Mo   & (NM) &  3.10 & 2.24 &  46.3 & -295.7  \\
Tc   & (NM) &  3.32 & 2.20 &  41.1 & -24.0  \\
     &      &       &   5$d$ metals  &     &    \\
 W   & (NM) & 3.00 & 2.31 & 49.6 & -127.8   \\
Re   & (NM) & 3.20 & 2.21 & 43.7 & -102.6   \\
Ir   & (FM) & 4.00 & 2.27 & 28.4 & 207.4   \\ 
Pt   & (NM) & 4.27 & 2.37 & 25.8 & 412.0   \\
Au   & (NM) & 4.60 & 2.55 & 28.2 & 330.6   \\
\end{tabular}
\end{ruledtabular}
\end{table}

The calculated equilibrium structural parameters (Fig. 1b), spin magnetic moment and magnetization energy
of the 4$d$ and 5$d$ TM zigzag chains are listed in Table IV. First of all, Table IV shows that all
the zigzag chains except the Re one, look like planar equilateral triangle ribbens, i.e.,
the two bond lengths $d_1$ and $d_2$ are similar and the angle $\alpha$ is close to 60$^{\circ}$
(Fig. 1b). The equilibrium bond lengths $d_1$ and $d_2$ are generally a few percents larger than
the bond length $d$ of the corresponding linear chains (Table I). This is because the zigzag chains which
form planar equilateral triangle ribbens, have a higher coordination number (four) than that (two) of
the linear chains. Similarly, all these bond lengths are shorter than their counterparts in the
bulk structures. For example, the bond lengths for bcc Nb, bcc Mo, fcc Rh, fcc Pd, bcc W, fcc Ir and
fcc Pt are, 2.86, 2.73, 2.68, 2.75, 2.86, 2.72 and 2.77 \AA, respectively.~\cite{kit96}

Our calculated equilibrium structural parameters ($d_1, d_2, \alpha$) agree reasonably well
with available previous calculations\cite{Ribeiro03,lin04,Lucas07}. For example, Lin {\it et al.} reported
$d_1 = 2.86$ \AA, $d_2 = 2.74$ \AA, $\alpha = 58.5^{\circ}$ for the Zr zigzag chain, being consistent with
our values in Table IV. Reported parameters $d_1$, $d_2$, and $\alpha$ for the W (2.44, 2.59, 61.9),
Os (2.48, 2.56, 61.1), Pt (2.58, 2.73, 61.9) and Au (2.64, 2.73, 61.13) chains
(estimated from Figs. 3 and 5 in Ref. \onlinecite{Lucas07}) are in rather good agreement
with our results in Table IV. One exception is the Ir chain\cite{Lucas07} where
$d_1 = 2.50$ \AA, $d_2 = 4.53$ \AA, $\alpha = 74.0^{\circ}$ differs substantially from the present results.
Secondly, all the 4$d$ and 5$d$ TM zigzag chains except that of Nb, Tc, La, Hf, Ta and Pt, have
magnetic solutions in the equilibrium structures (Table IV).
Further, the Zr, Mo, Ru, Rh, W, Re, Os, and Ir zigzag chains are
most stable in the FM state, whilst the ground state of the Pd zigzag chain is antiferromagnetic.
For comparison, the ground state of the linear Mo, Tc, W and Re chain is antiferromagnetic
(Table I).
The FM Ru and Rh zigzag chains have a rather large spin moment of $\sim$1.5  $\mu_B$/atom,
though the other magnetic zigzag chains generally have a small spin moment
($\leq$ 1.0  $\mu_B$/atom) (Table IV).
Interestingly, the ground state of the Y zigzag chain is nonmagnetic,
though it has a FM solution with
a spin magnetic moment of $\sim$0.5 $\mu_B$/atom.
Note that none of the 4$d$ and 5$d$ TMs is magnetic in their bulk structures in nature.
Thirdly, for some 4$d$ and 5$d$ transition metals, the
ground state magnetic configuration changes when the structure changes from
the linear to zigzag chain. For example, the Tc and Hf elements
are nonmagnetic in their equilibrium zigzag chain
structures, though they are, respectively, antiferromagnetic and ferromagnetic
in their equilibrium linear chain structures (Table I).
This is due to the increase in the coordination number in the zigzag chains because most of
them form a planar equilateral triangle ribben. On the other hand, the Y and Os elements become
ferromagnetic in the zigzag chains even though they are nonmagnetic in the linear chains.
Finally, the ground state of the Mo, W and Re chains changes from the AF state in the linear
chain to the FM state in the zigzag structure.

The Zr\cite{lin04}, Ru\cite{Ribeiro03}, Os, Au\cite{Lucas07}, Ir and Pt \cite{Lucas07,Victor09} 
zigzag chains were reported to have a metastable non-triangular elongated zigzag structure 
with $\alpha$ being $\sim 30.0^{\circ}$. The existence of this second energy minimum elongated 
zigzag structure ($\alpha$ being $\sim 30.0^{\circ}$) is believed to be crucial to the 
formation of a transient atomic chain in the break-junction experiments.\cite{Lucas07,Victor09}
To systematically study these possible elongated zigzag structures, we therefore further calculated
the total energy as a function of the fixed lattice constant $d_1$ with
$d_1$ varying from 2.0 \AA$ $ to 6.0 \AA$ $ for all the 4$d$ and 5$d$ zigzag chains.  
The structural parameters for the second local (or global) energy minimum state
of the zigzag chains are listed in Table V.  
Note that the equilibrium structural parameters listed in Table IV were obtained
by unrestricted structural relaxations using the conjugate gradient method (see Sec. II).
Our present calculations corroborate some of these previous findings.
For example, in the metastable Zr, Ir, Pt and Au chains, we find the angle $\theta$ to be
29.6$^\circ$,28.4$^\circ$, 25.8$^\circ$, 28.2$^\circ$, $d_{1}$ = 4.25, 4.00, 4.27, 4.60 \AA, and
spin moment $m_s = 0.30, 0.28, 0.00, 0.00 \mu_B$/atom, respectively.
In these metastable Zr, Ir, Pt and Au zigzag chains, the total energy
is, respectively, 1.02, 0.21, 0.41, 0.33 eV/atom higher than the ground state triangular
zigzag chains.
However, we don't find a second local (or global) energy minimum state in the
Ru and Os zigzag chains, in contrast to the previous 
studies.\cite{Ribeiro03,Lucas07}
The discrepancy on the Ru chain between the present and previous\cite{Ribeiro03} studies
could be attributed to the fact that highly accurate PAW potential rather than norm-conserving
pseudopotential, is used here, while the difference on 
the Os chain 
might be caused by the use of the faster but less accurate norm-conserving pseudopptential
linear combination of atomic orbitals method in Ref. \onlinecite{Lucas07}.
Surprisingly, the second energy minimum state in the Nb, Mo, Tc, W and Re
chains (Table V) is in fact the global energy minimum, i.e., its total energy is below
the corresponding minimum energy listed in Table IV. 
This result for the W chain is in agreement with the previous study of Ref. \onlinecite{Lucas07}.
Moreover, the bondlength $d_1$ of this second minimum state is not much larger than
that of the first minimum state and the angle $\alpha$ is not close to $30.0^{\circ}$.
These results appear to be consistent with the observation that only Ir, Pt and Au
could form an atomic chain in the break-junction experiments.
It could be worthwhile to search for the atomic chains in the break-junction experiments using Zr.
Finally, the Ir and Pt zigzag chains were reported to have a high-spin to low-spin transition 
near the local energy minimum.\cite{Lucas07, Victor09} In the present studies, 
the spin magnetic moment for Ir (Zr) in the ladder-like structure
is 0.629 (0.162) $\mu_B$/atom but becomes 0.285 (0.295) $\mu_B$/atom in the elongated energy minimum structure. 
In the W and Re zigzag chains, we found a
magnetic to nonmagnetic transition from the first energy minimum to the second energy minimum. 
For the Nb, Mo, Tc, Pt and Au zigzag chains,
both first and second  energy minimum states are nonmagnetic, 
and therefore, no high-spin to low-spin transition occurs.

\begin{table}
\caption{Spin ($m_s$) and orbital ($m_o$) magnetic moments (in $\mu_B$/atom) of the magnetic
4$d$ and 5$d$ transition metal zigzag chains in the equilibrium structures (Table IV)
with magnetization parallel ({\bf m}$\parallel \hat{z}$) and perpendicular ({\bf m}$\parallel \hat{x}$,
{\bf m}$\parallel \hat{y}$) (see Fig. 1) to the chain axis from fully relativistic
charge self-consistent calculations. Superscript $a$ denotes the orbital moments on two neighboring
atoms are antiparallel, though the system is in spin ferromagnetic state.
Superscript $\hat{y}$ means that the orbital moment is along the $y$-axis, though the
spin moment is along the $x$-axis, i.e., the spin and orbital moments are noncollinear.}
\begin{ruledtabular}
\begin{tabular}{cccccccc}
   &      &\multicolumn{2}{c}{{\bf m}$\parallel \hat{z}$} &\multicolumn{2}{c}{{\bf m}$\parallel \hat{x}$}
          &\multicolumn{2}{c}{{\bf m}$\parallel \hat{y}$} \\
   &      & $m_s$  & $m_o$   & $m_s$ & $m_o$ & $m_s$ & $m_o$ \\ \hline
          \multicolumn{8}{c}{4{\it d} metals} \\
 Y & (FM) & 0.982  & 0.004  & 0.979 & 0.079$^{a,\hat{y}}$ & 0.981 & 0.079$^a$ \\
Zr & (FM) & 0.162  & -0.003  & 0.162 & 0.149$^{a,\hat{y}}$ & 0.162 & -0.002$^a$ \\
Tc & (AF) &        &         &       &        & 0.032 & -0.257 \\
Ru & (FM) & 1.526  &  0.151  & 1.379 &  0.104 & 1.452 &  0.030 \\
   & (AF) & 0.286  &  0.144  & 0.263 &  0.024 & 0.261 &  0.009 \\
Rh & (FM) & 1.356  &  0.338  & 1.321 &  0.205 & 1.321 &  0.085 \\
Pd & (FM) &        &         & 0.182 &  0.045 & 0.155 &  0.022 \\
   & (AF) & 0.226  &  0.018  & 0.237 &  0.070 & 0.229 &  0.049 \\
          \multicolumn{8}{c}{5{\it d} metals} \\
Ta & (AF) &        &         &       &        & 0.111 & -0.266 \\
W  & (FM) & 0.261  & -0.042  &       &        & 0.163 &  0.002 \\
Re & (FM) & 0.517  & -0.018  & 0.506 & 0.141$^{a,\hat{y}}$ & 0.504 &  0.145 \\
Os & (AF) &        &         &       &        & 0.095 & -0.160 \\
Ir & (FM) & 0.690  &  0.433  &       &        &       &        \\
   & (AF) &        &         &       &        & 0.157 &  0.458 \\
Pt & (AF) &        &         &       &        & 0.139 &  0.325 \\
Au & (AF) &        &         &       &        & 0.038 &  0.135 \\
\end{tabular}
\end{ruledtabular}
\end{table}

When the SOC is taken into account, not only the spin magnetic moments would depend on the magnetization
direction but also the orbital magnetic moments would appear. In the 3$d$ TM chains, the spin magnetic
moments are hardly affected by the SOC~\cite{Tung07} because of the smallness of the SOC in these systems.
In contrast, in the 5$d$ TM chains, the SOC is so large that it not only would affect the size of
the magnetic moments but also could suppress or induce magnetism itself, depending on the magnetization
orientation, as mentioned already in Sec. IIIb. The magnetic moments in the magnetic zigzag
chains for the magnetization along
three coordinate axes from fully relativistic charge self-consistent calculations are listed in Table V.
We notice that all 5$d$ TM chains exhibit the remarkable CMA\cite{smo08} behavior. Even two 4$d$ (AF Tc and
FM Pd) zigzag chains show the CMA too. In particular, in the FM Ir chain, the magnetism occurs only when
the magnetization is along the chain. In contrast, in the AF Ir chain, the magnetism appears only
when the magnetization is parallel to the $y$-axis (Fig. 1b).
The orbital magnetic moments in the Zr, Tc, Ru, Rh, Ta, Re, Os, Ir, Pt, Au chains are rather
significant ($\geq$ 0.1 $\mu_B$/atom) (Table V). In the Rh, Ir and Pt zigzag chains, the orbital
magnetic moments can be as large as 0.3 $\mu_B$/atom.
As in the linear chains (Table II), the orbital moments in the zigzag chains depend strongly
on the magnetization orientation (Table VI).

\subsection{Band structures and density of states}

\begin{figure}
\includegraphics[width=8cm]{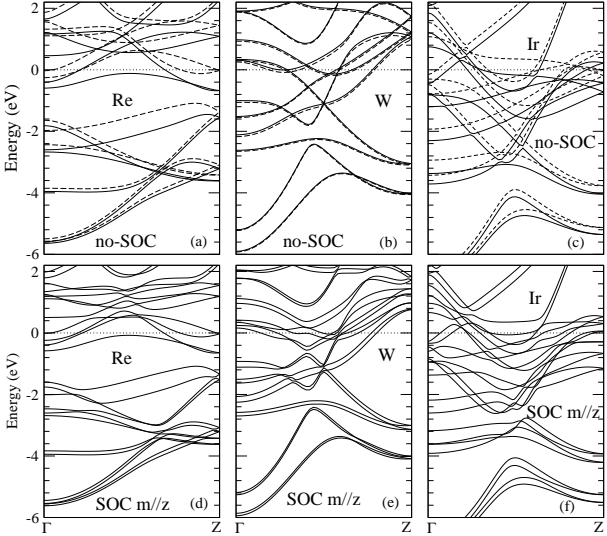}
\caption{Scalar-relativistic (a-c) and fully relativistic (d-f) band structures of the Re, W, and Ir
zigzag atomic chains in the FM state. In (d-f), the spin magnetization is along the
chain direction (i.e., the $z$-axis). The Fermi level (the dotted horizontal line)
is at the zero energy.}
\end{figure}

The band structures of the FM Re, W and Ir zigzag chains are
displayed in Fig. 8, as representatives. Compared with the corresponding band
structures of the linear chains (e.g., Ir in Fig. 5), the number of bands become doubled in the
zigzag chains because of the doubling of the number of atoms. Furthermore, unlike
the linear chains where the $d_{xy} (d_{xz})$ and $d_{x^2-y^2} (d_{yz})$ bands (Fig. 5d)
are degenerate because of rotational invariance, the $d_{xy} (d_{xz})$ and $d_{x^2-y^2} (d_{yz})$ bands
are now split because of the strong anisotropy in the $x-y$ plane perpendicular to the
chain axis. It is clear that the energy bands are also highly spin-split and the separation
of the spin-up and spin-down bands may be correlated with the spin magnetic moment.

As for the linear chains, we calculate the spin-polarization ($P$) and count the numbers of spin-up
and spin-down conduction bands at the Fermi level in the FM zigzag chains, as listed in Table III.
The $P$ in the considered zigzag chains generally gets reduced
when compared with that in the linear chains (Table III). Nevertheless,
the $P$ of the Rh zigzag chain is as large as 0.53. Interestingly, the sign of the
$P$ in the Zr chain changes from positive to negative when it transforms
from the linear to zigzag chain structure.

\section{Stability of linear chain structures}

Let us now examine the relative stability of the linear and zigzag chains by
comparison of the total energies of the two structures.
The ground state cohesive
energy of the linear chains and the cohesive energies of the
zigzag chains in the NM, FM and AF states are displayed in Fig. 9 (for 4$d$ TM) and Fig. 10.
(for 5$d$ TM). The cohesive energy ($E_c$) is defined as the difference in the total energy
between the free atom ($E_{a}$) and the chain ($E_{t}$), i.e. $E_c = E_{a} - E_{t}$.
A positive value of the $E_c$ means that the formation of the chain from the
free atoms would save energy, i.e., the chain would be stable against breaking up
into free atoms. The total energies of the free atoms are calculated by the cubic
box supercell approach with the cell size of 15 \AA. The electronic configurations
used for 4$d$ TM are $4d^15s^2$ (Y), $4d^35s^1$ (Zr), $4d^45s^1$ (Nb), $4d^55s^1$ (Mo),
$4d^65s^1$ (Tc), $4d^75s^1$ (Ru), $4d^85s^1$ (Rh), $4d^95s^1$ (Pd) and $4d^{10}5s^1$ (Ag).
And for 5$d$ TM are $5d^16s^2$ (La), $5d^36s^1$ (Hf), $5d^46s^1$ (Ta), $5d^56s^1$ (W),
$5d^66s^1$ (Re), $5d^76s^1$ (Os), $5d^86s^1$ (Ir), $5d^96s^1$ (Pt) and $5d^{10}6s^1$ (Au).

\begin{figure}
\includegraphics[width=7cm]{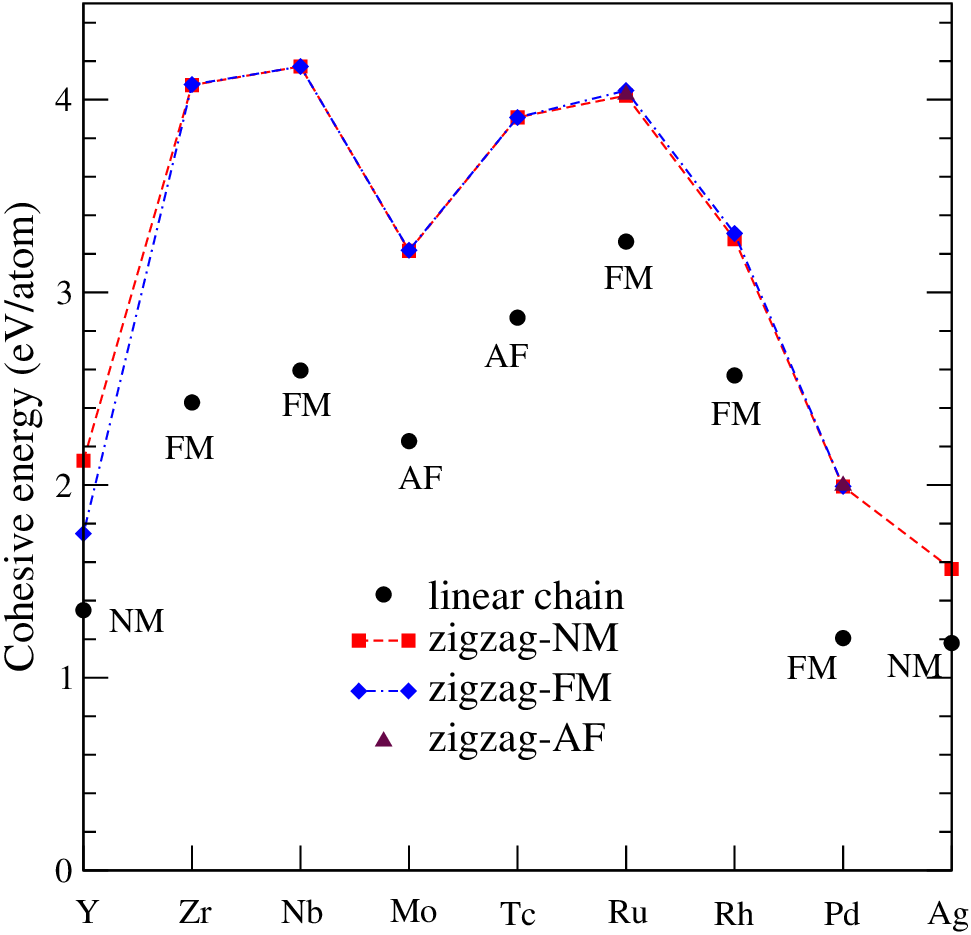}
\caption{(color online) The cohesive energy of the 4$d$ TM zigzag chains in the
NM, FM and AF states. For comparison, the ground state cohesive energy of
the corresponding linear chains is also plotted (solid circles).
The ground state magnetic configuration of the linear chains is labelled
as NM or FM or AF near each solid circle.}
\end{figure}

\begin{figure}
\includegraphics[width=7cm]{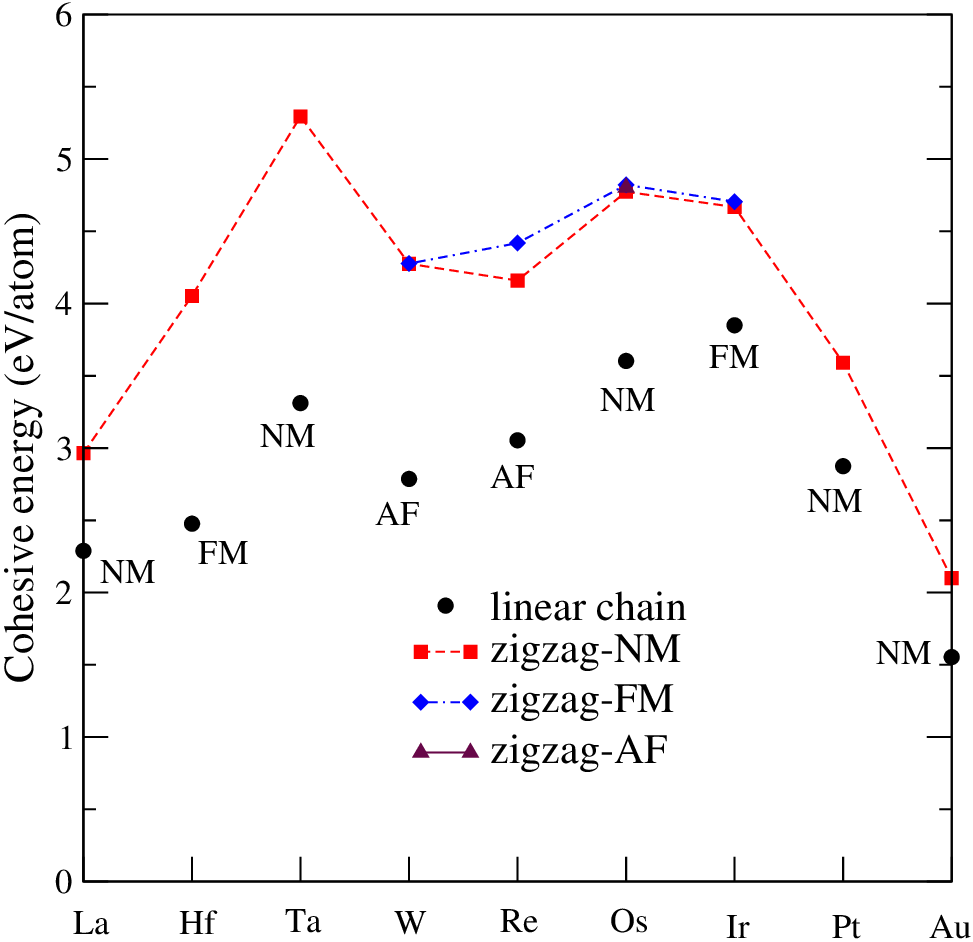}
\caption{(color online) The cohesive energy of the 5$d$ TM zigzag chains in the
NM, FM and AF states. For comparison, the ground state cohesive energy of
the corresponding linear chains is also plotted (solid circles).
The ground state magnetic configuration of the linear chains is labelled
as NM or FM or AF near each solid circle.}
\end{figure}

We note that in all cases, the ground state cohesive energy
of the linear chain is smaller than that of the zigzag chain (Figs. 9 and 10).
This suggests that the 4$d$ and 5$d$ linear chains are unstable against the zigzag structural
distortion, as may be expected from the Peierls instability of linear one-dimensional monoatomic
metals.~\cite{kit96}
The difference in the ground state energy between the linear and zigzag structures
for all the 4$d$ and 5$d$ elements is rather large, ranging from 0.8 to 2.0 eV/atom.
This shows that the free standing 4$d$ and 5$d$ TM linear chains would not be the stable state,
and the linear chains may occur only in constrained conditions such as on the
steps on a vicinal surface~\cite{Gambar02} and under tensile stress in
the break-point experiments~\cite{Rodrigues03,Rodrigues00,Jiandong06,Susumu06}.
Interestingly, a recent {\it ab initio} study\cite{cho07} showed that alloying the
gold nanowires with cesium could make linear monoatomic chains stable.

\section{Magnetic Anisotropy Energy}

The total energy as a function of the magnetization orientation ($\theta,\phi$)
of a 1D wire may be written, in the lowest non-vanishing terms, as
\begin{equation}
E_t=E_{0}+sin^{2}\theta (E_{1}-E_{2}cos^{2}\phi)
\end{equation}
where $\theta$ is the polar angle of the magnetization away from the chain axis ($z$-axis)
and $\phi$ is the azimuthal angle in the $x-y$ plane perpendicular to the wire, measured from the
{\it x} axis. For the free standing linear atomic chains, the azimuthal anisotropy
energy constant $E_{2}$ is zero. The axial anisotropy energy
constant $E_1$ is then given by the total energy difference between the magnetization along
the $y$($x$) and {\it z} axes, i.e., $E_1 = E^y - E^z$ ($E^x = E^y$). A positive value of $E_1$
means that the chain ($z$) axis is the easy magnetization axis. For the zigzag chains which are
in the $x-z$ plane, $E_{2}$ is not zero and can be calculated as the total energy difference
between the magnetization along the $x$ and $y$ axes, i.e., $E_2 = E^y - E^x$.

\begin{table}
\caption{Total ($E_1^t$), electronic ($E_1^e$) and dipolar ($E_1^d$) magnetic anisotropy energies
(in meV/atom) of the 4$d$ and 5$d$ transition metal linear chains.
If $E_1^t$ is positive, the easy magnetization axis is along the chain; otherwise, the easy
magnetization axis is perpendicular to the chain.}
\begin{ruledtabular}
\begin{tabular}{ccccccc}
   & \multicolumn{3}{c}{FM} & \multicolumn{3}{c}{AF} \\
   & $E_1^t$ & $E_1^e$ & $E_1^d$ & $E_1^t$ & $E_1^e$ & $E_1^d$ \\ \hline
      &        &      & 4$d$ metals    &     &  &              \\
Zr & -0.277 &  -0.286 & 0.009 &     &          &        \\
Mo &        &         &       &  -2.783  &  -2.924  &  0.141 \\
Tc &        &         &       &   7.228  &   7.186  &  0.042 \\
Ru &-11.99 & -12.03 & 0.044 &     &          &        \\
Rh &  6.997 &   6.993 & 0.004 &     &          &        \\
Pd & -1.760 &  -1.770 & 0.012 &     &          &        \\ \hline
      &        &      & 5$d$ metals    &     &       &         \\
Hf &  0.825 &   0.825 & 0.000 &     &          &        \\
W  &        &         &       &  -5.235  &  -5.283  &  0.048 \\
Re &        &         &       &  -59.94  &  -60.01   &  0.070 \\
Ir &-11.13 & -11.14 & 0.014 &     &          &        \\
\end{tabular}
\end{ruledtabular}
\end{table}

  The magnetic anisotropy energy for a magnetic solid consists of two contributions.
One comes from the magnetocrystalline anisotropy in the electronic band structure
caused by the simultaneous occurrence of the electron spin-orbit interaction
and spin-polarization in the magnetic system, and
{\it ab initio} calculation of this part has already been described in Sec. II.
The other is the magnetostatic (or shape) anisotropy energy due to the magnetic dipolar
interaction in the solid.
The shape anisotropy energy is zero for the cubic systems such as bcc Fe and fcc Ni,
and also negligibly small for weakly anisotropic solids such as hcp Co.
However, for the highly anisotropic structures such as magnetic Fe and Co monolayers,~\cite{Guo,Guo2}
the shape anisotropy energy can be comparable to the electronic MAE, and therefore cannot be
neglected.
For the collinear magnetic systems ({\it{i.e.}} {\bf{m$_{q}$}}//{\bf{m$_{q^{'}}$}}),
this magnetic dipolar energy {\it $E_{d}$} is given by (in atomic Rydberg units)~\cite{Guo}
\begin{equation}
 E^{d} = \sum_{qq^{'}}{\frac{m_{q}m_{q^{'}}}{c^2} M_{qq^{'}}}
\end{equation}
where  $M_{qq^{'}}$ is called the magnetic dipolar Madelung constant which is evaluated
by Ewald's lattice summation technique \cite{Ewald}. The speed of light $c$ = 274.072,
and {$m_{q}$} is the atomic magnetic moment on site {\it q} in the unit cell. Note that in atomic Rydberg units,
one Bohr magneton ($\mu_B$) is $\sqrt{2}$. Therefore, as noted recently in Ref. \onlinecite{Tung07},
the $E_{d}$ for the multilayers obtained previously by Guo {\it et al.}\cite{Guo,Guo2} is too small by a
factor of 2.

The calculated $E^d$'s for the linear and zigzag chains
are listed in Tables VII and VIII, respectively. Tables VII and VIII show that in both the linear
and zigzag chains and in both the FM and AF states, the $E^d$'s are much smaller
than the electronic contributions ($E^e$), being in strong contrast to the case of the 3$d$ TM chains.\cite{Tung07}
This is because the magnetization here is significantly lower and the equilibrium bond length
becomes larger, compared with that of the 3$d$ TM chains.\cite{Tung07}
Furthermore, they always prefer the chain direction ($z$ axis) as the easy magnetization axis.
Therefore, any perpendicular magnetic anisotropy must originate from the
electronic magnetocrystalline anisotropy.

The calculated $E^e$'s of the linear and zigzag atomic chains are also
listed in Tables VII and VIII, respectively. Table VII shows that in the FM linear
chains at equilibrium, the $E^e$ would favor a perpendicular anisotropy
in the Zr, Ru, Pd and Ir chains but prefer the chain axis in the Rh and Hf chains.
In the AF state,
in contrast, the Mo, W and Re linear chains would have the easy axis perpendicular
to the chain while only the Tc linear chain prefer the axial anisotropy.
Remarkably, the FM Ru, Rh and Ir as well as AF Tc and Re linear chains
have a large total anisotropy energy ($E^t$) (see Table VII) of $\sim$10 meV/atom.
In particular, the $E^t$ of the AF Re linear chain is as large as -60 meV/atom.
{\it Ab initio} calculations of the $E^e$ of the 4$d$ TM linear chains have been
reported recently\cite{Mokrousov06}, and our present results for the equilibrium
bondlengths (Table VII) agree rather well with these earlier calculations (Fig. 1 in
Ref. \onlinecite{Mokrousov06}).

The electronic anisotropy energy for the selected linear 4$d$ and 5$d$
atomic chains is displayed as a function of bond length in Fig. 11.
It is clear that in several selected linear chains, the magnitude of the $E^e$
generally increases with the bondlength (Fig. 11a), like the spin and orbital
magnetic moments (Fig. 4). For example, the $E^e$ of the Rh chain increases
from 7.0 meV/atom at the equilibrium bondlength (2.25 \AA) to 37.3 meV/atom
at bondlength of 3.0 \AA. Several chains also undergo interesting spin-reorientation
transition as the bondlength is elongated.

When elongated, for example, the FM Zr, Nb, Ru, and Ir linear chain would undergo a spin
reorientation transition from the perpendicular to along the axial direction at
the bondlength of $\sim$2.75\AA, $\sim$2.82 \AA, $\sim$2.65\AA, and $\sim$2.45\AA, respectively.
In contrast, the magnetization of the Ta chain transits from that along the axis to the
perpendicular direction at $\sim$2.85\AA.
Furthermore, many elongated chains have a gigantic anisotropy energy of $\sim$20 meV/atom
(Fig. 11).

Table VIII shows that the size of the axial anisotropy energy ($E_1$) in the
zigzag chains is large and is generally comparable to that in the linear chains
(Table VII). However, unlike the linear chains, there is also the pronounced
anisotropy ($E_2$) in the $x-y$ plane perpendicular to the chain axis in
many zigzag structures (Table VIII). In the FM Y, AF Ru and Os zigzag chains,
the magnitude of the $E^t_2$ is even larger than that of $E^t_1$.
In the FM Y, Mo, AF Ru, FM Rh, FM Re, and FM Ir zigzag chains,
the easy axis is along the chain direction. In the FM Ru as well as FM and
AF Pd chains, the easy axis is perpendicular to the zigzag plane.
In the Zr, W, and Os chains,
the easy axis is in the zigzag plane but perpendicular to the chain axis.
{\it Ab initio} calculations for only the Ir and Pt zigzag chains have recently
been reported.\cite{Lucas07} However, in Ref. \onlinecite{Lucas07}, the easy axis
is reported to be along the $x$-axis. This discrepancy may be due to the
pronounced difference in the equilibrium zigzag structure between the present
and previous calculations. Furthermore, we find the Pt zigzag chain
to nonmagnetic.

Structural transformation from the linear to zigzag structure has profound
effect on magnetism in the 4$d$ and 5$d$ TM nanowires.
This transformation not only induces (or suppresses) magnetization in, e.g.,
the Y and Os chains (the Tc and Hf chains), as mentioned already in Sec. IV,
but also causes spin reorientation transition in, e.g., the Re and Ir chains
(Tables VII and VIII). Note that the linear AF Re chain has
a gigantic perpendicular anisotropy energy of -60.0 meV/atom (Table VII).
However, upon transition to the zigzag structure, the AF state
disappears, and, instead, the FM state appears with the magnetization
switched to be along the chain axis.

\begin{table*}
\caption{The total ($E_1^t$, $E_2^t$), electronic ($E_1^e$, $E_2^e$) and dipolar ($E_1^d$, $E_2^d$) magnetic
anisotropy energy constants (in meV/atom) as well as the easy magnetization axis
({\bf M}) of the 4$d$ and 5$d$ transition metal zigzag chains.
$E_1$ = $E^y$ - $E^z$; $E_2$ = $E^y$ - $E^x$, see Equ. (2).}
\begin{ruledtabular}
\begin{tabular}{ccccccccccccccc}
   & \multicolumn{7}{c}{FM} &\multicolumn{7}{c}{AF}  \\
   &$E_1^e$ &$E_2^e$ &$E_1^d$ &$E_2^d$ & $E_1^t$ &$E_2^t$ & {\bf M}&$E_1^e$ &$E_2^e$ &$E_1^d$ &$E_2^d$ & $E_1^t$ &$E_2^t$ & {\bf M}\\ \hline
    &       &        &       &       &      &       & 4$d$ metals&  &   &      &      &     &     &   \\
Y   & 0.028 & -1.925 & 0.004 & 0.002 & 0.032&-1.923 & $z$ &      &      &      &      &     &     &   \\
Zr  & 0.000 &  0.018 & 0.001 & 0.000 & 0.001& 0.018 & $x$ &      &      &      &      &     &     &   \\
Mo  &0.009&0.009& 0.001&0.000 &0.010 &0.009& $z$ &  & &    &      &     &     &  \\
Ru  &-2.673&-0.084&0.084 & 0.041& -2.589& -0.043& $y$ &0.441&-0.544& 0.004 & -0.002 & 0.445   & -0.548 & $z$ \\
Rh  &10.675&3.182&0.059 &0.029 &10.734 & 3.211& $z$&      &      &      &      &     &     &  \\
Pd  &-0.695&-0.487& 0.005& 0.002& -0.690& -0.485& $y$ &-1.125&-0.860& 0.001 & -0.001  & -1.123& -0.861& $y$\\ \hline
    &       &        &       &       &      &       & 5$d$ metals&  &   &      &      &     &     &   \\
W   &-0.540&0.240& 0.002& 0.001& -0.538& 0.241& $x$ &      &      &      &      &     &     &  \\
Re  & 1.062&1.043& 0.008& 0.004&  1.070 &1.047 & $z$ &      &      &      &      &     &     &  \\
Os  &-9.402&0.319& 0.007& 0.004&-9.395 & 0.323& $x$ &-4.798&5.340& 0.002 & -0.002 & -4.796 &5.338  & $x$\\
Ir  &17.595&-6.430&0.014 &0.006 &17.609 &-6.424 & $z $ &      &      &      &      &     &     &  \\
\end{tabular}
\end{ruledtabular}
\end{table*}

\begin{figure}
\includegraphics[width=8cm]{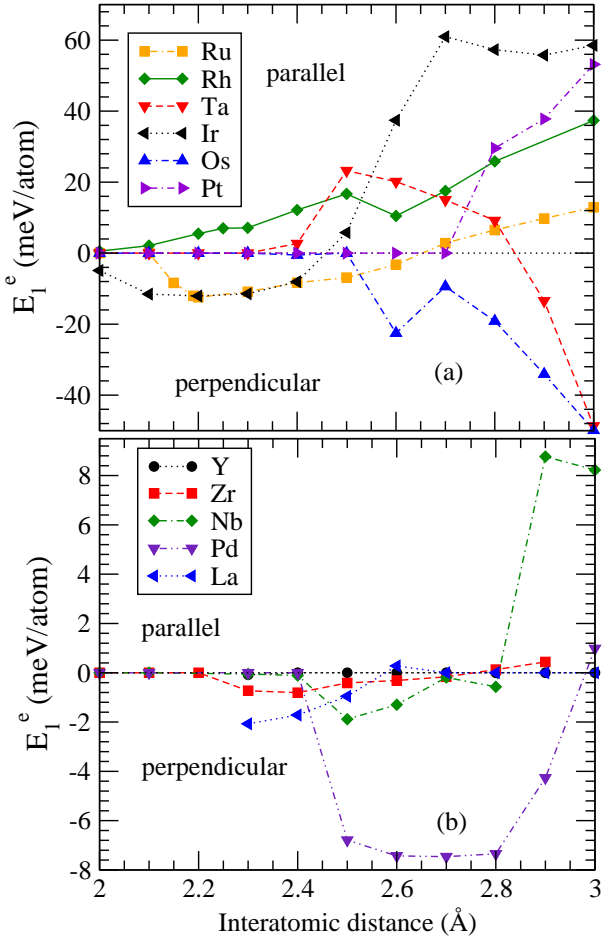}
\caption{(color online) Magnetocrystalline anisotropy energy ($E_1^e$) of the selected
4$d$ and 5$d$ transition metal linear atomic chain as a function of interatomic distance.
The upper panels contains the TM linear chain with larger MAE. A positive value of $E_1^e$
means that the magnetization would be parallel to the chain axis whilst a negative value
would means that the easy magnetization axis would be perpendicular to the chain.}
\end{figure}

\section{Conclusions}

We have performed an extensive {\it ab initio} study of the physical
properties of both linear and zigzag atomic chains of all 4$d$ and 5$d$
transition metals within the GGA by using the accurate PAW method. First, the atomic structures
were determined. All the TM linear chains are found to be unstable against the
corresponding zigzag structures. All the TM chains except Nb, Ag and La, have a stable (or metastable)
magnetic state in either the linear or zigzag or both structures. Magnetic states appear also
in the Nb and La linear chains when the chains are sufficiently elongated. The spin magnetic
moments in the Mo, Tc, Ru, Rh, W, Re chains could be large ($\geq$1.0 $\mu_B$/atom).
Structural transformation from the linear to zigzag chains can suppress the magnetism already
in the linear chain, induce the magnetism in the zigzag structure, and also cause a change
of the magnetic state (FM to AF or vice verse).

With the SOC included, our calculations show that the orbital moments in the Zr, Tc, Ru, Rh, Pd,
Hf, Ta, W, Re, Os, Ir and Pt chains could be rather large ($\geq$0.1 $\mu_B$/atom).
Importantly, large magnetic anisotropy energy ($\geq$1.0 meV/atom) is found in
most of the magnetic TM chains, suggesting that these nanowires could have important
applications in ultrahigh density magnetic memories and hard disks.
In particular, giant magnetic anisotropy energy ($\geq$10.0 meV/atom) could appear in the
Ru, Re, Rh, and Ir chains. Furthermore, the magnetic anisotropy energy in several
linear chains could be as large as 40.0 meV/atom when the chains are under sufficiently
large tensile strain. A spin-reorientation transition occurs in the Ru, Ir, Ta, Zr, La and
Zr, Ru, La, Ta and Ir linear chains when they are elongated.
Remarkably, all the 5$d$ as well as Tc and Pd chains show the fascinating behavior of
the so-called colossal magnetic anisotropy.\cite{smo08}
Finally, the electronic band structure and density of states of the nanowires
have also been calculated mainly in order to understand the electronic origin
of the large magnetic anisotropy and orbital magnetic moment as well as to
calculate the conduction electron spin polarization.

\section{Acknowledgements}
The authors acknowledge supports from National Science Council and NCTS of Taiwan.
They also thank National Center for High-performance Computing of Taiwan and NTU Computer
and Information Networking Center for providing CPU time.

\end{document}